\documentclass[prl,preprint,superscriptaddress]{revtex4-1}
\usepackage[space]{grffile}
\usepackage{bm,graphicx,amsmath,amssymb,color}
\usepackage{euscript,multirow,tabularx}
\usepackage{longtable}
\usepackage{float}
\usepackage{calrsfs}
\usepackage{amsmath}
\usepackage[draft]{todonotes}
\usepackage{fancyvrb}
\usepackage[utf8]{inputenc}
\usepackage{todonotes}

\graphicspath{{/home/klein/misc/share/latex/fmrfm/yig/phonons/interfer/figs/png/}}

\begin{document}

\title{Coherent long-range transfer of angular momentum between magnon Kittel
  modes by phonons}


\author{K. An}
\affiliation{Univ. Grenoble Alpes, CEA, CNRS, Grenoble INP,
  Spintec, 38054 Grenoble, France}

\author{A.N. Litvinenko}
\affiliation{Univ. Grenoble Alpes, CEA, CNRS, Grenoble INP, Spintec, 38054 Grenoble, France}

\author{R. Kohno}
\affiliation{Univ. Grenoble Alpes, CEA, CNRS, Grenoble INP, Spintec, 38054 Grenoble, France}

\author{A.A. Fuad}
\affiliation{Univ. Grenoble Alpes, CEA, CNRS, Grenoble INP, Spintec, 38054 Grenoble, France}

\author{V. V. Naletov} 
\affiliation{Univ. Grenoble Alpes, CEA, CNRS, Grenoble INP, Spintec, 38054 Grenoble, France}
\affiliation{Institute of Physics, Kazan Federal University, Kazan
    420008, Russian Federation}

\author{L. Vila}
\affiliation{Univ. Grenoble Alpes, CEA, CNRS, Grenoble INP, Spintec, 38054 Grenoble, France}

\author{U. Ebels}
\affiliation{Univ. Grenoble Alpes, CEA, CNRS, Grenoble INP, Spintec, 38054 Grenoble, France}

\author{G. de Loubens} 
\affiliation{SPEC, CEA-Saclay, CNRS, Universit\'e Paris-Saclay,
  91191 Gif-sur-Yvette, France}

\author{H. Hurdequint} 
\affiliation{SPEC, CEA-Saclay, CNRS, Universit\'e Paris-Saclay,
  91191 Gif-sur-Yvette, France}

\author{N. Beaulieu} 
\affiliation{LabSTICC, CNRS, Universit\'e de Bretagne Occidentale,
  29238 Brest, France}

\author{J. Ben Youssef} 
\affiliation{LabSTICC, CNRS, Universit\'e de Bretagne Occidentale,
  29238 Brest, France}

\author{N. Vukadinovic}
\affiliation{Dassault Aviation, Saint-Cloud 92552, France} 

\author{G.E.W. Bauer} 
\affiliation{Institute for Materials Research and WPI-AIMR and CSRN, Tohoku University, Sendai 980-8577, Japan}

\author{A.~N. Slavin} \affiliation{Department of Physics, Oakland
  University, Michigan 48309, USA}

\author{V.~S. Tiberkevich} \affiliation{Department of Physics, Oakland
  University, Michigan 48309, USA}

\author{O. Klein}
\email[Corresponding author:]{ oklein@cea.fr}
\affiliation{Univ. Grenoble Alpes, CEA, CNRS, Grenoble INP,
  Spintec, 38054 Grenoble, France}

\date{\today}

\begin{abstract}
  We report ferromagnetic resonance in the normal configuration of an
  electrically insulating magnetic bi-layer consisting of two yttrium iron
  garnet (YIG) films epitaxially grown on both sides of a 0.5 mm thick
  non-magnetic gadolinium gallium garnet (GGG) slab. An interference pattern is
  observed and it is explained as the strong coupling of the magnetization
  dynamics of the two YIG layers either in-phase or out-of-phase by the standing
  transverse sound waves, which are excited through the magneto-elastic
  interaction. This coherent mediation of angular momentum by circularly
  polarized phonons through a non-magnetic material over macroscopic distances
  can be useful for future information technologies.
\end{abstract}

\maketitle

The renewed interest in using acoustic oscillators as coherent signal
transducers \cite{Bienfait2019, Moores2018, Tsaturyan2017} stems from the
extreme finesse of acoustic signal transmission lines. The low sound attenuation
factor $\eta_{a}$ benefits the interconversion process into other wave forms
(with damping $\eta_{s}$) as measured by the cooperativity,
$\mathcal{C}=\Omega^{2}/(2 \eta_{a}\eta_{s})$ \cite{Al-Sumaidae2018,
  Spethmann2015}, leading to strong coupling as defined by $\mathcal{C}>1$ even
when the coupling strength $\Omega$ is small. Here we present experimental
evidence for coherent long-distance transport of angular momentum via the
coupling to circularly polarized sound waves that exceeds previous benchmarks
set by magnon diffusion \cite{Cornelissen2015, Oyanagi2019, Lebrun2018} by
orders of magnitude.

The material of choice for magnonics is yttrium iron garnet (YIG) with the
lowest magnetic damping reported so far \cite{spencer59, cherepanov93}. The
ultrasonic attenuation coefficient in garnets is also exceptional, \textit{i.e.}
up to an order of magnitude lower than that in single crystalline quartz
\cite{LeCraw1965, Spencer1962}. Spin-waves (magnons) hybridize with lattice
vibrations (phonons) by the magnetic anisotropy and strain dependence of the
magneto-crystalline energy \cite{Kittel1958, Boemmel1959, Damon1965a, Seavey1965,
  Dreher2012, Zhang2016}. Although often weak in absolute terms, the
magneto-elasticity leads to new hybrid quasiparticles (``magnon polarons'') when
spin-wave (SW) and acoustic-wave (AW) dispersions (anti)cross \cite{gurevich96,
  Doetsch1978, Wago1998a}. This coupling has been exploited in the past to
produce microwave acoustic transducers \cite{Pomerantz1961, Reeder1969},
parametric acoustic oscillators \cite{Chowdhury2015} or nonreciprocal acoustic
wave rotation \cite{Popa2014, Matthews1962}. Recent studies have identified
their beneficial effects on spin transport in thin YIG films by pump-and-probe
Kerr microscopy \cite{Ogawa2015,Hashimoto2017} and in the spin Seebeck effect
\cite{Kikkawa2016}. The adiabatic conversion between magnons and phonons in
magnetic field gradients proves their strong coupling in YIG \cite{Holanda2018}.

\begin{figure}
  \includegraphics[width=0.7\textwidth]{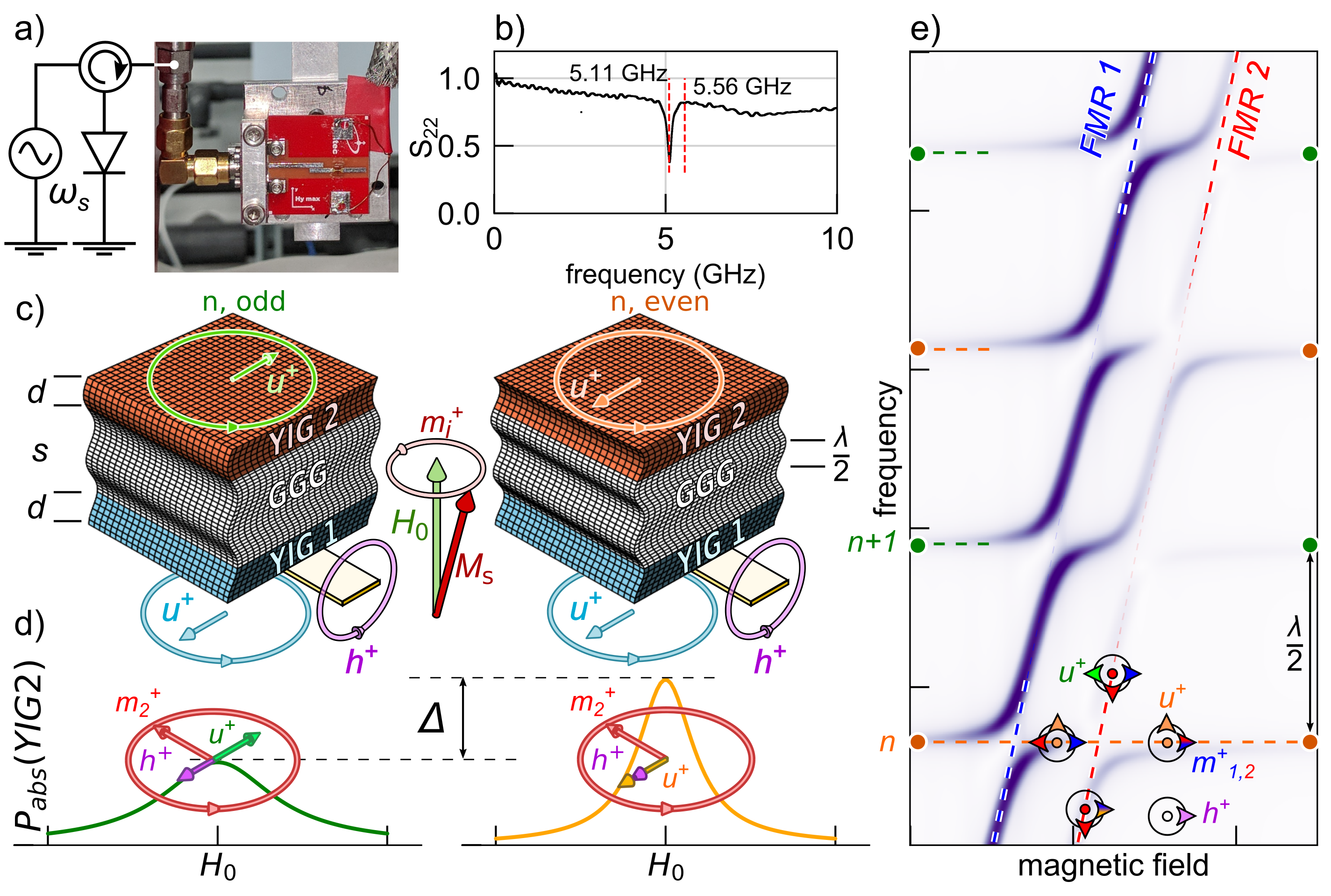}
  \caption{(Color online) a) Schematic and picture of the ferromagnetic
    resonance (FMR) setup. A butterfly shaped stripline resonator \cite{Yap2013}
    with 0.3~mm wide constriction is in contact with the bottom layer of the
    YIG1($d=200$~nm)$\vert$GGG($s=0.5$~mm)$\vert$YIG2($d=200$~nm) ``dielectric
    spin-valve'' stack. The microwave antenna can be tuned in or out of its
    fundamental resonance ($5.11$~GHz) as shown in the reflectivity spectrum b).
    c) Schematic of the coupling between the top (red) and bottom (blue) YIG
    layers by the exchange of coherent phonons: the magnetic precession $m^{+}$
    generates a circular shear deformation $u^{+}$ of the lattice that can be
    tuned into a coherent motion of all fields. Constructive/destructive
    interference between the dynamics of the two YIG layers occurs for even/odd
    mode numbers $n$ causing d) a contrast $\Delta$ in the absorbed microwave
    power ($P_\text{abs}$) between tones separated by half a phonon wavelength.
    e) Density plot of the spectral modulation of $P_\text{abs}$ produced by
    Eq.(\ref{eq:coupling}) when magnetic bi-layers are strongly coupled to
    coherent phonon modes. The orange/green dots indicate the spectral position
    of the even/odd acoustic resonances.}
  \label{FIG1}
\end{figure}

But phonons excited by magnetization dynamics can also transfer their angular
momentum into an adjacent non-magnetic dielectrics \cite{Comstock1963a,
  Garanin2015}. When the latter acts as a phonon sink, the ``phonon pumping''
increases the magnetic damping \cite{Streib2018}. The substrate of choice for
YIG is single crystal gadolinium gallium garnet (GGG) which in itself has very
long phonon mean-free path \cite{Spencer1963, Dutoit1972} and small impedance
mismatch with YIG \cite{Polzikova2019}, raising the hope of a phonon-mediated
dynamic exchange of coherence through a non-magnetic insulating layer
\cite{Streib2018}.

Here we report ferromagnetic resonance experiments (FMR) of a ``dielectric
spin-valve'' stack consisting of half a millimeter thick single-crystal GGG slab
coated on both sides by thin YIG films. We demonstrate that GGG can be an
excellent conductor of phononic angular momentum currents allowing the coherent
coupling between the two magnets over millimeter distance. Figure~\ref{FIG1}a
illustrates the experimental setup in which an inductive antenna monitors the
coherent part of the magnetization dynamics. The spectroscopic signature of the
dynamic coupling between the two YIG layers is a resonant contrast pattern as a
function of microwave frequency (see intensity modulation along FMR2 in
Figure~\ref{FIG1}e).

Before turning to the experimental details, we sketch a simple phenomenological
model that captures the dynamics of the fields as described by the continuum
model for magneto-elasticity with proper boundary conditions \cite{Streib2018}.
The perpendicular dynamics of a trilayer with in-plane translational symmetry
can be mapped on three coupled harmonic oscillators, viz. the Kittel modes of
the two magnetic layers $m_{i=1,2}$ and the $n$-th mechanical mode, $u_n$, in
the dielectric, which obey the coupled set of equations
\begin{subequations}
\begin{align}
  (\omega_s -\omega_1 + j \eta_s)  m_1^+ & = \Omega_1 u_n^+/2 + \kappa_1 h^+\\
  (\omega_s -\omega_2 + j \eta_s)  m_2^+ & = \Omega_2 u_n^+/2 + \kappa_2 h^+\\
  (\omega_s -\omega_{n} + j \eta_a)\: u_n^+ & = \Omega_1 m_1^+/2 +
  \Omega_2 m_2^+/2
\end{align} \label{eq:coupling}
\end{subequations}
Here $\omega _{n}/(2\pi) =v/ \lambda_{n}$, where $v$ is the AW velocity and
$\lambda_{n}/2 = (2d+s)/n$ is a half wavelength that fits into the total sample
thickness$\ 2d+s$, with $n$ being an integer (mode number). The dynamic
quantities $m_{i}^{+}=\left( m_{x}+jm_{y}\right) _{i}$ are circularly polarized
magnetic complex amplitudes ($j$ being the imaginary unit) precessing
anti-clockwise around the equilibrium magnetization at Kittel resonance
frequencies $\omega _{1}\neq \omega _{2}$. In our notation $\eta _{s/a}$ are the
magnetic/acoustic relaxation rates \cite{Rueckriegel2014} and the constants
$\Omega _{i}$ and $\kappa _{i}$ are the magneto-elastic interaction and
inductive coupling to the antenna, respectively. Coherence effects between
$m_{1}$ and $m_{2}$ can be monitored by the power $P_{\mathrm{abs}}=\kappa
_{i}\text{Im} (h^{\star }m_{i})$ as a function of the microwave frequency
$\omega _{s}$ of the driving field with circular amplitude $h^+$ \footnote{the
  antenna produces a linear rf field, which decomposes in both a left and right
  circulating field with only one component coupling to the magnetization
  dynamics}. Note that Eq.(\ref{eq:coupling}) holds when the characteristic AW
decay length exceeds the film thickness (see below).

The acoustic modes with odd and even symmetry couple with opposite signs, i.e.
$\Omega _{2}=(-1)^{n}\Omega _{1}$ (see Figure~\ref{FIG1}c), which affects the
dynamics as sketched in Figure~\ref{FIG1}d. When $n$ is odd (even), the top
layer returns (absorbs) the power from the electromagnetic field, because the
phonon amplitude is out-of(in) phase with the direct excitation, corresponding
to destructive (constructive) interference. In other words, the phonons pumped
by the dynamics of the layer 1 are reflected vs. absorbed by layer 2. According
to Eq.(\ref{eq:coupling}), a contrast $\Delta$ should emerge between tones
separated by half a wavelength. This is illustrated in Figure~\ref{FIG1}e by
plotting the calculated modulation of the magnetic absorption when two Kittel
modes with slightly different resonance frequencies and different inductive
coupling to the antenna interact via strong coupling to coherent phonons (see
below values in Table.(\ref{tab:mat})). In the figure the effect is more visible
around the resonance of the layer with weaker coupling $\kappa_{2}<\kappa_{1}$
to the antenna (FMR2, red dashed line) since, according to the model, the
amplitude of the contrast is proportional to the amplitude ratio of the
microwave magnetic fields felt by the two YIG layers: $\Delta \propto \kappa
_{1}/\kappa _{2}$. We employ here a stripline with width ($0.3$~mm) that couples
strongly to the lower layer YIG1, while still allowing to monitor the FMR
absorption of YIG2.\footnote{We disregard the inhomogeneity in the driving field
  generated by the local antenna.}

Figure~\ref{FIG1}a is a picture of the bow-tie $\lambda /2$-resonator (with
reflectivity spectrum shown in Figure~\ref{FIG1}b) with which we perform
spectroscopy around $5$~GHz. The later fulfills the ``half-wave condition'' of
the phonon relative to the YIG thickness that maximizes the phonon pumping
\cite{Streib2018}. The sample was grown by liquid phase epitaxy, i.e. by
immersing a GGG monocrystal substrate with thickness $s=0.5$~mm and orientation
(111) into molten YIG. The concomitant growth leads to nominally identical YIG
layers, with thickness $d=200$~nm on both sides of the GGG. The Gilbert damping
parameter $\alpha \approx 9\times 10^{-5}$, measured as the slope of the
frequency dependence of the line width, is evidence for the high crystal
quality. All experiments have been carried out at room temperature and on the
same sample. Because of that, the results shall be presented in inverse
chronological order.

\begin{figure}
  \includegraphics[width=0.7\textwidth]{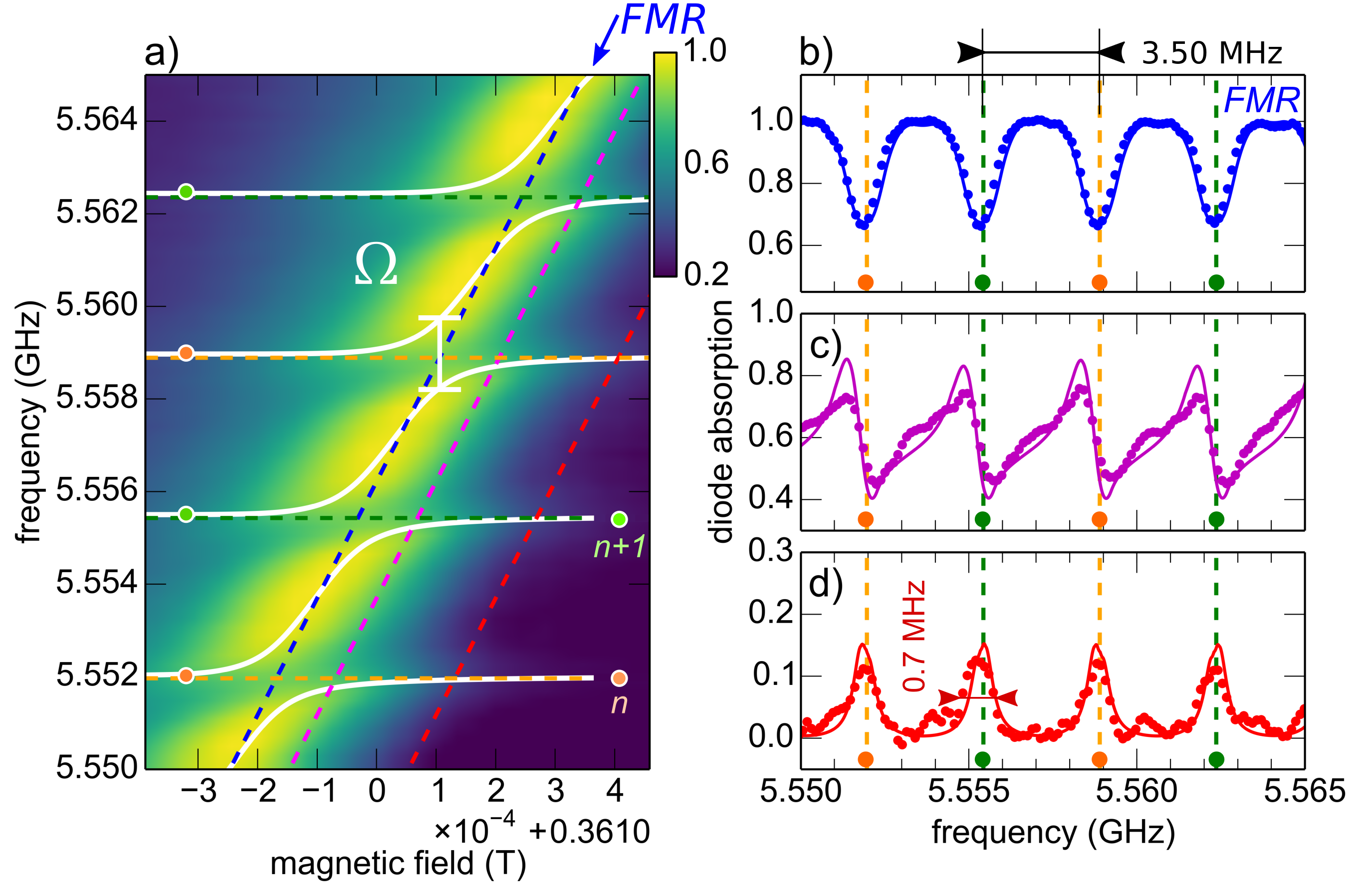}
  \caption{(Color online)a) Microwave absorption spectra of a
    YIG(200~nm)$\vert$GGG(0.5~mm) crystal, revealing a periodic modulation of
    the intensity interpreted as the avoided-crossing between the FMR mode (see
    blue arrow) at $\omega_1 = \gamma \mu_0 (H_0 -M_1)$, and the $n^\text{th}$
    standing (shear) AW$_\text{n}$ resonances across the total thickness
    (horizontal dash lines in orange and green) at $\omega_n = n \pi v /(d+s)$ .
    The right panels (b,c,d) show the intensity modulation for 3 different cuts
    (blue, magenta and red) along the gyromagnetic ratio (\textit{i.e.} parallel
    to the resonance condition). The solid lines in the 4 panels are fits by the
    oscillator model (cf. Eq.(\ref{eq:coupling}) with fit values in
    Table.(\ref{tab:mat})).}
  \label{FIG2}
\end{figure}

Having removed YIG2 by mechanical polishing, we first concentrate on the dynamic
behavior of a single magnetic layer. Figure~\ref{FIG2}a shows the FMR absorption
of YIG1$\vert$GGG bilayer \cite{Gulyaev1981, Ye1988, Litvinenko2015,
  Tikhonov2017} around $5.56$~GHz \textit{i.e.} for a detuned antenna having
weak inductive coupling. These spectra are acquired in the perpendicular
configuration, where the magnetic precession is circular, by magnetizing the
sample with a sufficiently strong external magnetic field, $H_0$, applied along
the normal of the films. Figure~\ref{FIG2}a provides a detailed view of the fine
structure within the FMR absorption that is obtained when one sweeps the
field/frequency in tiny steps of 0.01~mT/0.1~MHz, respectively.

The FMR mode (see arrow) follows the Kittel equation $\omega_{1} \approx \gamma
\mu_0 (H_0- M_{1})$ \footnote{The exact expression is more complicated and
  contains cubic/uniaxial anistropies of
  $H_\text{k1}=-7.8$~mT/$H_\text{ku}=-3.75$~mT respectively.}, with $\gamma/(2
\pi)=28.5$~GHz/T, the gyromagnetic ratio and $\mu_0 M_1=0.1720$~T, the
saturation magnetization, but its intensity vs. frequency is periodically
modulated \cite{Ye1988, Ye1991} which we explain by the hybridization with the
comb of standing shear AWs described by Eq.(\ref{eq:coupling}) truncated to one
magnetic layer.

We ascribe the periodicity of 3.50~MHz in the signal of Figures~\ref{FIG2} to
the equidistant splitting of standing phonon modes governed by the
\emph{transverse} sound velocity of GGG along (111) of $v=3.53\times 10^{3}$~m/s
\cite{Ye1988, Ye1991, Khivintsev2018} via $v/(2d+2s)\approx 3.53$~MHz
\footnote{the total crystal thickness reduces to $d+s$ after polishing}. This
value thus separates two phononic tones, which differ by half a wavelength. At
5.5~GHz, the intercept between the transverse AW and SW dispersion relations
occurs at $2\pi/\lambda_n = \omega_s /v \approx 10^{5}$~cm$^{-1}$, which
corresponds to a phonon wavelength of about $\lambda_n\approx 700$~nm with index
number $n\sim 1400$. The modulation is strong evidence for the high acoustic
quality that allows elastic waves to propagate coherently with a decay length
exceeding twice the film thickness, \textit{i.e.} 1~mm. For later reference we
point out that the absorption is the same for odd and even phonon modes, whose
eigen-values are indicated here by green and orange dots.

In Figures~\ref{FIG2}bcd we focus on the line shapes at detunings parallel to
the FMR resonance as a function of field and frequency indicated by the blue,
magenta, and red cuts in Figure~\ref{FIG2}a. The amplitude of the main resonance
(blue line) in Figure~\ref{FIG2}b dips and the lines broaden at the phonon
frequencies \cite{Ye1988, Ye1991}. The minima transform via dispersive-looking
signal (magenta in \ref{FIG2}ac) into peaks (red \ref{FIG2}ad) once sufficiently
far from the Kittel resonance as expected from the complex impedance of two
detuned resonant circuits, illustrating a constant phase between $m_i$ and $u_n$
along these cuts. The $m_i$ are circularly polarized fields rotating in the
gyromagnetic direction, that interact only with acoustic waves $u_n$ with the
same polarity, as implemented in Eq.~(\ref{eq:coupling})
\cite{Holanda2018}.

The observed line shapes can be used to extract the lifetime parameters in
Eq.~(\ref{eq:coupling}). We first concentrate on the observed $0.7$~MHz full
line width of the acoustic resonances in Figure~\ref{FIG2}d. Far from the Kittel
condition, the absorbed power is governed by the sound attenuation. According to
Eq.~(\ref{eq:coupling}), the absorbed power at large detuning reduces to
$P_{\text{abs}} \propto ((\omega_{s}-\omega_{n})^{2} + \eta_{a}^{2})^{-1}$. The
AW decay rate $\eta_{a}/(2\pi)=0.35$~MHz is obtained as the half line width of
the acoustic resonance, leading to a characteristic decay length $\delta =
v/\eta_{a}\approx2$~mm for AW excited around 5.5~GHz. The acoustic amplitude
therefore decays by $\sim 20\%$ over the half millimeter film thickness. The
sound amplitude in both magnetic layers are therefore roughly the same, as
assumed in Eq.(\ref{eq:coupling}). This figure is consistent with the measured
ultrasonic attenuation in GGG: 0.70~dB/$\mathrm{ \mu }$s at 1GHz
\cite{Dutoit1972, Dutoit1974}, \textit{i.e.}, a lifetime of about
0.5~$\mathrm{\mu }$s at 5GHz.

The SW lifetime $1/\eta_{s}$ follows from the broadening of the absorbed power
at the Kittel condition which contains a constant inhomogeneous contribution and
a frequency-dependent viscous damping term. When plotted as function of
frequency, the former is the extrapolation of the line widths to zero frequency,
in our case $\sim$ 5.7~MHz (or 0.2~mT). On the other hand, the Gilbert
phenomenology (see above) of the homogeneous broadening
$\eta_{s}=\alpha\omega_{s}$ corresponds to a $\eta _{s}/(2\pi)=0.50$~MHz at
5.5~GHz. The dominantly inhomogeneous broadening is here caused by thickness
variations, a spatially dependent magnetic anisotropy, but also by the
inhomogeneous microwave field.

Conspicuous features in Figure~\ref{FIG2}a are the clearly resolved
avoided-crossing of SW and AW dispersion relations, which prove the strong
coupling between two oscillators. Fitting by hand the dispersions of two coupled
oscillators through the data points (white lines), we extract a gap of
$\Omega/(2\pi)=1$~MHz and a large cooperativity $\mathcal{C}\approx3$. From the
overlap integral between a standing shear AW confined in a layer of thickness
$\sim s$ and the Kittel mode confined in a layer of thickness $d$, one can
derive the analytical expression for the magneto-elastic coupling strength
\cite{khymyn2019, Ye1988}:
\begin{equation}
  \Omega = \dfrac B {\sqrt{2}}\, \sqrt {
    \dfrac {\gamma } {\omega_s M_1 \, \rho s d}} 
  \left ( 1 - \cos \omega_s \dfrac d v \right ) \label{eq:omega}
\end{equation}
where \cite{Spencer1963} $B=(B_2+2B_1)/3=7\times 10^5$~J/m$^3$, with $B_1$ and
$B_2$ being the magneto-elastic coupling constants for a cubic crystal, and
$\rho=5.1$~g/cm$^3$ is the mass density of YIG. From Eq.(\ref{eq:omega}) we
infer that coherent SW excited around $\omega_s/(2\pi)\approx5.5$~GHz have a
dynamic coupling to shear AW of the order of $\Omega /(2\pi )=1.5$~MHz, close
to the value extracted from the experiments.

The material parameters extracted for our YIG$\vert$GGG are summarized in
Table~(\ref{tab:mat}). Numerical solutions of Eq.~(\ref{eq:coupling}) using
these values are shown as solid lines in Figure~\ref{FIG2}bcd. The agreement
with the data is excellent, confirming the validity of the model and parameters.
\begin{table}
  \caption{Material parameters used in the oscillator model (all values are
    expressed in units of $2\pi\times 10^6$~rad/s). }
  \begin{ruledtabular}
    \begin{tabular}{*{5}{c}}
      {${\omega_{1}-\omega_{2}} $} &
      {${\omega_{n+1}-\omega_{n}} $} &
      {${\Omega} $} &
      {${\eta_s} $} & 
      {${\eta_a} $} \\
      \hline
      {40}  & 
      {3.50}  & 
      {1.0} &
      {0.50}  & 
      {0.35}
\end{tabular}
\end{ruledtabular}\label{tab:mat}
\end{table}

\begin{figure}
  \includegraphics[width=0.7\textwidth]{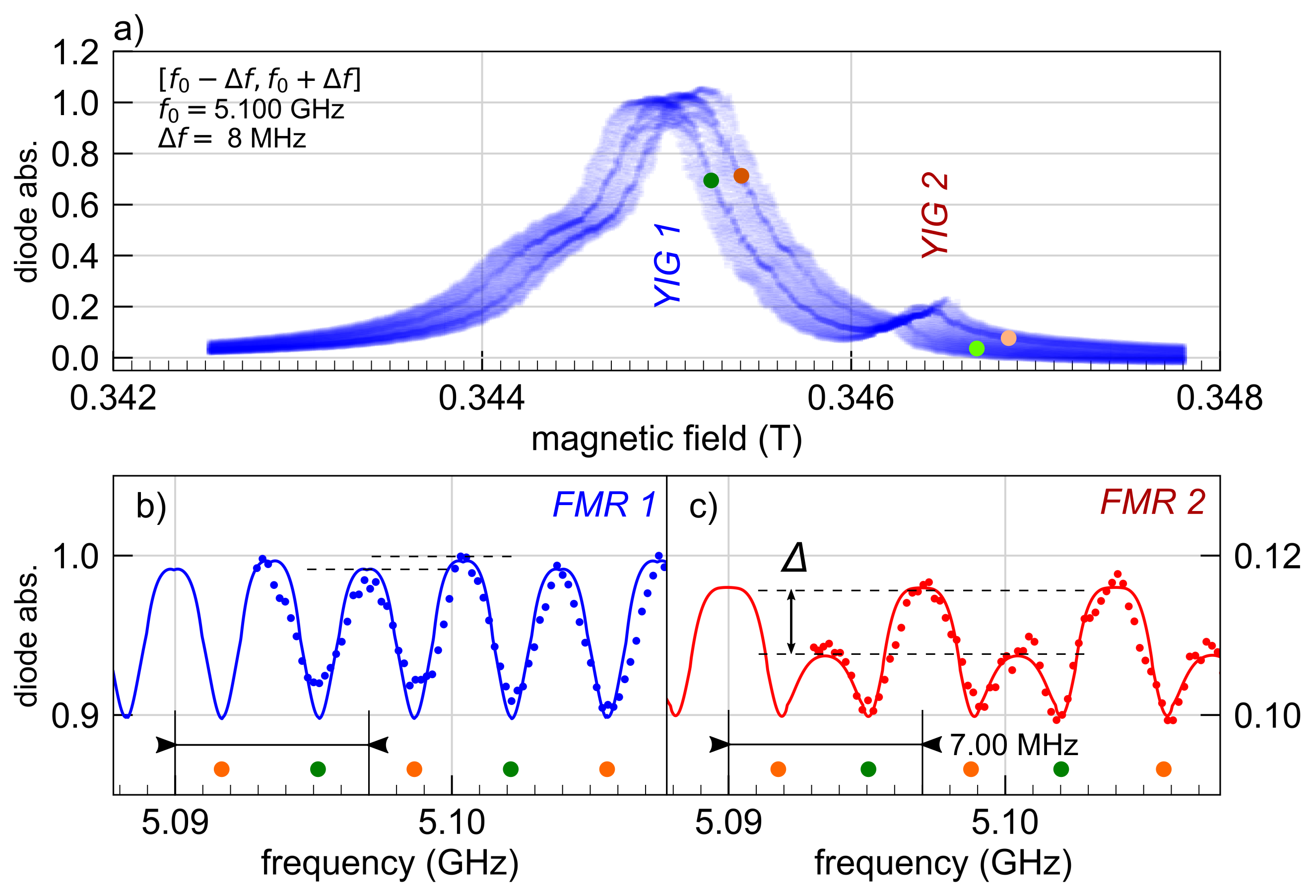}
  \caption{(Color online) FMR spectroscopy of the YIG1$\vert$GGG$\vert$YIG2
    trilayer. Panel a) is a transparent overlay of magnetic field sweeps for
    frequencies in the interval $5.101\pm 0.008$~GHz by 0.1~MHz steps. Dark
    lines reveal two acoustic resonances marked by orange and green dots. Panel
    b) and c) show the frequency modulation of the FMR amplitude for
    respectively the bottom YIG1 layer and the top YIG2 layer, in which a
    contrast $\Delta $ appears between neighboring acoustic resonances. The
    solid lines show the modulation predicted by Eq.(\ref{eq:coupling}).}
  \label{FIG3}
\end{figure}

The other needed parameter for solving Eq.(\ref{eq:coupling}) in the general
case is the attenuation ratio $\kappa_{2}/\kappa_{1}\approx7$ deducted from a
factor of 50 decreased power when flipping the single YIG layer sample upside
down on the antenna. The layer is then separated 0.5~mm from the antenna, and
the observed reduction agrees with numerical simulations using electromagnetic
field solvers.

We turn now our attention to the magnetic sandwich in which YIG1 touches the
antenna and the nominally identical YIG2 is 0.5~mm away, where a slight
difference in uniaxial anisotropy causes separate resonance frequencies. Since
we want to detect also the resonance of the top layer, we have to compensate for
the decrease in inductive coupling by tuning the source frequency to the antenna
resonance at 5.11~GHz (see Figure~\ref{FIG1}b). This enhances the signal by the
quality factor $Q\sim30$ of the cavity at the cost of an increased radiative
damping of the bottom layer signal \cite{Bloembergen1954}.

Figure~\ref{FIG3}a is a transparent overlay of field sweeps for frequency steps
of 0.1~MHz in the interval $5.101\pm 0.008$~GHz. We attribute the two peaks
separated by 1.4~mT (or 40~MHz) to the bottom and top YIG Kittel resonances, the
later shifted due to a slight difference in effective magnetization
$\mu_{0}M_{2}= \mu_{0}M_{1}+0.0014$~T. Note that the detuning between the two
Kittel modes is large compared to the strength of the magneto-elastic coupling
$\Omega$. In Figure~\ref{FIG3}b and Figure~\ref{FIG3}c we compare the measured
modulation of the resonance amplitude for respectively the bottom YIG1 layer and
top YIG2 layers. This corresponds to performing 2 cuts at the resonance
condition FMR1 and FMR2 in the same fashion as Figure~\ref{FIG2}b. The top YIG2
signal is modulated with a period of 7.00~MHz (Figure~\ref{FIG3}c) with a
contrast $\Delta$ between even and odd modes. This agrees with the prediction of
Eq.(\ref{eq:coupling}) (see solid lines) due to constructive/destructive
couplings mediated by even/odd phonon modes, the modulation period of the
absorbed power doubles along the resonance of the top layer (FMR2), when
compared to the case of a single YIG layer (Figure~\ref{FIG2}).
Figure~\ref{FIG3}b illustrates also that the strong coupling $\kappa_{1}$ to the
antenna hinders clear observation of this modulation in the bottom YIG1 layer
resonance. Nevertheless, the anticipated sign change of $\Delta$ (by the
inverted phase of $u_n$ relative to $m_2$ in Eq.(\ref{eq:coupling})) between
FMR1 and FMR2 remains observable.

\begin{figure}
  \includegraphics[width=0.7\textwidth]{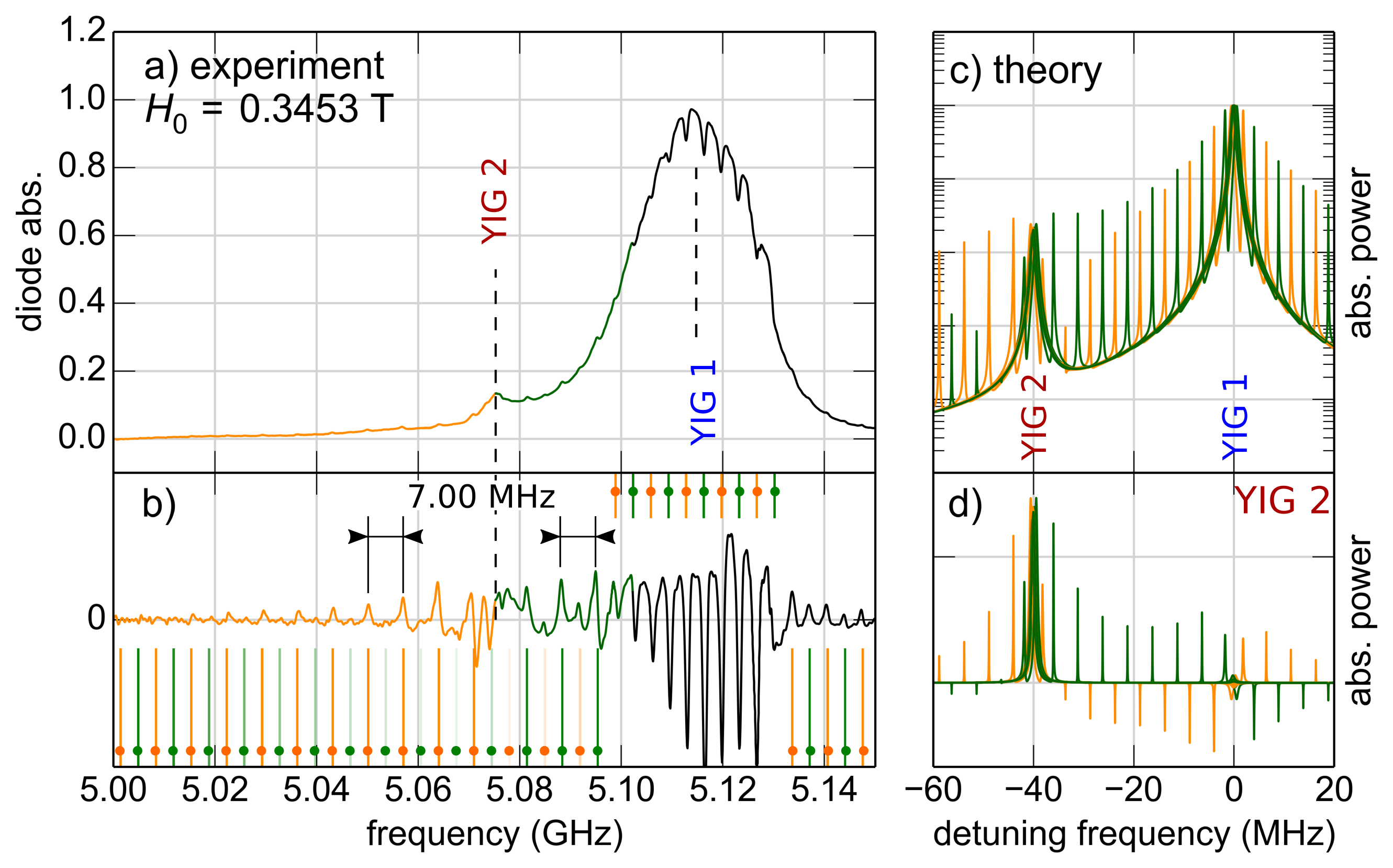}
  \caption{(Color online) a) Frequency sweep at fixed field performed on the
    magnetic bi-layer. The fine regular modulation within the FMR envelop is
    ascribed to the excitation of acoustic shear waves resonances. The acoustic
    pattern is enhanced in panel b) by subtracting the FMR envelop emphasizing
    the constructive/destructive interferences of the even/odd acoustic
    resonances in the vicinity of the YIG2 FMR mode. Panel c) shows on a
    logarithmic scale the predicted modulation using the experimental parameters
    of Table.(\ref{tab:mat}). Panel d) shows on a linear scale the corresponding
    power absorbed by the top magnetic layer only.}
  \label{FIG4}
\end{figure}

We now address the acoustic resonances revealed by the dark lines in
Figure~\ref{FIG3}a for odd/even indices labeled by green/orange circles in the
wings. The phonon line with even index (orange marker) progressively disappears
when approaching the YIG2 Kittel resonance from the low field (left side) of the
resonance, while the opposite behavior is observed for the odd index feature
(green marker), which disappears when approaching the YIG2 Kittel resonance from
the high field (right side). This behavior agrees with the model in
Figure~\ref{FIG1}e. The contrast in the acoustic resonance intensity mirrors the
contrast of the amplitude of the FMR resonance.

Figure~\ref{FIG4}a shows the observed FMR absorption spectrum around 5.11~GHz
measured at fixed field $H_{0}=0.3453$~T. We enhance the fine structure in
Figure~\ref{FIG4}b by subtracting the FMR envelope and progressively amplifying
the weak signals in the wings. The orange/green color code emphasizes the
constructive/destructive interference of the even/odd acoustic resonances in the
top-layer signal. This feature can be explained by Eq.~(\ref{eq:coupling}), as
shown by the calculated curves in Figure~\ref{FIG4}cd. The acoustic modes change
character from even to odd (or vice versa) across the FMR frequency, which is
caused by the associated phase shift by 180$^{\circ}$ of the acoustic drive,
again explaining the experiments. The absorption by the YIG2 top layer in
Figure~\ref{FIG4}d may even become negative so the phonon current from YIG1
drives the magnetization in YIG2. This establishes both angular momentum and
power transfer of microwave radiation via phonons.

In summary, we report interferences between the Kittel resonances of two
ferromagnets over macroscopic distance through the exchange of circularly
polarized coherent shear waves propagating in a nonmagnetic dielectric. We show
that magnets are a source and detector for phononic angular momentum currents
and that these currents provide a coupling, analogous to the dynamic coupling in
metallic spin valves \cite{Heinrich2003}, but with an insulating spacer, over
much larger distances, and in the ballistic/coherent rather than
diffuse/dissipative regime. This should lead to the creation of a dynamical gap
between collective states when the two Kittel resonances are tuned within the
strength of the magneto-elastic coupling. Our findings might have implications
on the non-local spin transport experiments \cite{Cornelissen2017}, in which
phonons provide a parallel channel for the transport of angular momentum. While
the present experiments are carried out at room temperature and interpreted
classically, the high acoustic quality of phonon transport and the strong
coupling to the magnetic order in insulators may be useful for quantum
communication.

This work was supported in part by the Grants No.18-CE24-0021 from the ANR of
France, No. EFMA-1641989 and No. ECCS-1708982 from the NSF of the USA, by the
Oakland University Foundation, the NWO and Grants-in-Aid of the Japan Society of
the Promotion of Science (Grant 19H006450). V.V.N. acknowledges support from UGA
through the invited Prof. program and from the Russian Competitive Growth of
KFU. We would like to thank Simon Streib for illuminating discussions.


%

\end{document}